\documentstyle[prl,aps,epsf,multicol]{revtex}
\begin{document}
\draft
\newcommand{\npa}[1]{Nucl.~Phys.~{A#1}}
\newcommand{\etal}{{\em et al.}}
\title{Near-threshold production of the multi-strange $\Xi^-$ hyperon}
\author{ P.~Chung$^{(1)}$, N.~N.~Ajitanand$^{(1)}$, J.~M.~Alexander$^{(1)}$, 
M.~Anderson$^{(6)}$, D.~Best$^{(5)}$,
F.P.~Brady$^{(6)}$, T.~Case$^{(5)}$, W.~Caskey$^{(6)}$, D.~Cebra$^{(6)}$,
J.L.~Chance$^{(6)}$, B.~Cole$^{(11)}$, K.~Crowe$^{(5)}$,
A.~C.~Das$^{(3)}$, J.E.~Draper$^{(6)}$, M.L.~Gilkes$^{(1)}$,
S.~Gushue$^{(1,9)}$, M.~Heffner$^{(6)}$,
A.S.~Hirsch$^{(7)}$, E.L.~Hjort$^{(7)}$, W.~Holzmann$^{(1)}$, L.~Huo$^{(13)}$,
M.~Issah$^{(1)}$,M.~Justice$^{(4)}$,
M.~Kaplan$^{(8)}$, D.~Keane$^{(4)}$, J.C.~Kintner$^{(12)}$, J.~Klay$^{(6)}$,
D.~Krofcheck$^{(10)}$, R.~A. ~Lacey$^{(1)}$, J.~Lauret$^{(1)}$, 
M.A.~Lisa$^{(3)}$, H.~Liu$^{(4)}$, Y.M.~Liu$^{(13)}$,
J.~Milan$^{(1)}$, R.~McGrath$^{(1)}$, Z.~Milosevich$^{(8)}$,
G.~Odyniec$^{(5)}$, D.L.~Olson$^{(5)}$,
S.~Panitkin$^{(4)}$, N.T.~Porile$^{(7)}
$, G.~Rai$^{(5)}$, H.G.~Ritter$^{(5)}$,
J.L.~Romero$^{(6)}$, R.~Scharenberg$^{(7)}$, B.~Srivastava$^{(7)}$, 
N.T.B~Stone$^{(5)}$, T.J.M.~Symons$^{(5)}$, A.~Taranenko$^{(1)}$,
J.~Whitfield$^{(8)}$, T.~Wienold$^{(5)}$, R.~Witt$^{(4)}$, L.~Wood$^{(6)}$,
W.N.~Zhang$^{(13)}$
                        \\  (E895 Collaboration) \\
and\\
H.~Oeschler$^{(1,2)}$}
\address{ $^{(1)}$Depts. of Chemistry and Physics,
SUNY at Stony Brook, New York 11794-3400 \\
$^{(2)}$Darmstadt University of Technology, 64289 Darmstadt, Germany \\
$^{(3)}$Ohio State University, Columbus, Ohio 43210\\
$^{(4)}$Kent State University, Kent, Ohio 44242 \\
$^{(5)}$Lawrence Berkeley National Laboratory,Berkeley, California, 94720\\
$^{(6)}$University of California, Davis, California, 95616 \\
$^{(7)}$Purdue University, West Lafayette, Indiana, 47907-1396 \\
$^{(8)}$Carnegie Mellon University, Pittsburgh, Pennsylvania 15213\\
$^{(9)}$Brookhaven National Laboratory, Upton, New York 11973 \\
$^{(10)}$University of Auckland, Auckland, New Zealand \\
$^{(11)}$Columbia University, New York, New York 10027 \\
$^{(12)}$St. Mary's College, Moraga, California  94575 \\
$^{(13)}$Harbin Institute of Technology, Harbin, 150001 P.~R. China \\
}
\date{\today}
\maketitle
%
%
\begin{abstract}
%

The yield for the multi-strange $\Xi^{-}$ hyperon  has been measured in 6
AGeV Au+Au collisions via reconstruction of its decay products
$\pi^{-}$ and $\Lambda$, the latter also being reconstructed from its
daughter tracks of $\pi^{-}$ and p.  The measurement is rather close to the
threshold for $\Xi^{-}$ production and therefore provides an important
test of model predictions. The measured yield for $\Xi^{-}$ and $\Lambda$ are 
compared for several centralities. In central collisions the $\Xi^{-}$ yield
is found to be in excellent agreement with statistical and 
transport model predictions, suggesting that multi-strange hadron production
approaches chemical equilibrium in high baryon density nuclear matter.
\end{abstract}
\pacs{PACS 25.75.Dw}
\begin{multicols}{2}
\narrowtext

Many years ago, it was proposed that strange particle yields could serve
as an indicator for the formation of the quark-gluon plasma (QGP)\cite{Raf80}.
For a partonic system with  partially restored chiral symmetry, strangeness
is expected to be readily produced due to the lower energy threshold for
$s\bar{s}$ quark pair production. The production of strangeness may be
further increased for significant baryon densities where the chemical potential 
associated with the production of light quarks is raised. For these reasons 
it is expected that the QGP phase should reflect a high strange
(anti)quark density\cite{raf91}. Due to their strange quark content,
multi-strange baryons [and anti-baryons] are expected to be a more 
sensitive probe of this phase than hadrons which contain only one 
strange valence quark.

Interest in strangeness and especially in the yield of multi-strange hadrons 
has grown enormously\cite{sqm2001}.
The observation of ``strangeness enhancement'' in the production
of $\Xi^-$ ($dss$) and $\Omega^-$ ($sss$) in heavy ion collisions
compared to pp or pA collisions has contributed to the excitement \cite{WA97}.

	A comparison of experimental and calculated yields indicates two distinct
features:
(i)~Statistical models, with input of temperature and chemical potential,  
are in fair or good agreement with the yields reported from the 
Alternating Gradient Synchrotron (AGS)~\cite{PBM_AGS}, the Super Proton 
Synchrotron (SPS)~\cite{PBM_SPS} and from recent experiments performed at 
the Relativistic Heavy Ion Collider (RHIC)~\cite{PBM_RHIC,QM2002}.
Using a canonical description this approach is successful even 
for beam energies of $\sim$~1~AGeV~\cite{Clemans_01}. They can even 
explain the observed enhancement between pA and Pb+Pb
collisions\cite{Tounsi}.
(ii)~Transport-model calculations based on hadronic interactions are
currently unable to account for the measured yields of multi-strange
particles at SPS energies (158 AGeV)\cite{Soff}.
The success of the statistical model is surprising and raises the question
as to whether or not this success holds true for near-threshold multi-strange
particle production.

	This letter reports on the production of the singly-strange $\Lambda$ 
hyperon and the doubly-strange $\Xi^-$ hyperon in 6~AGeV Au+Au
collisions. Preliminary reports on $\Lambda$ hyperon production have 
been given elsewhere \cite{PChung98}. The $\Xi^-$ measurement reported here
is unique in that it represents the 
lowest incident energy for which such a measurement has been made in
heavy ion collisions (the laboratory energy threshold for direct production in $NN$
collisions is 3.74 GeV). The production of multi-strange hadrons is arguably most
interesting close to the production threshold due to the suppression of 
multistep processes (the $\Xi^-$ particle is  mainly produced from 
the strangeness-exchange reactions (${\bar K (\Lambda,\Sigma)\to\Xi (\pi,\eta)}$). 
Collisions of 6 AGeV Au~+~Au are expected to produce nuclear matter with high 
baryon density and relatively low chemical freeze-out temperature. 
Thus, marked changes in the model predictions are expected for multi-strange 
particle yields.

	The experiment has been performed with the EOS TPC\cite{grai90} at 
the Alternating Gradient Synchrotron (AGS) at Brookhaven National 
Laboratory (BNL). Details on the detector and its
setup have been reported earlier\cite{PChung98,grai90,CPinkenburg99}. 
The data presented here benefit from the excellent coverage,
continuous 3D-tracking, and particle identification capabilities of the TPC.
These features are crucial for the efficient detection and reconstruction 
of $\Lambda$'s and $\Xi^-$.

\begin{figure}

\centerline{\epsfysize=5.2in \epsffile{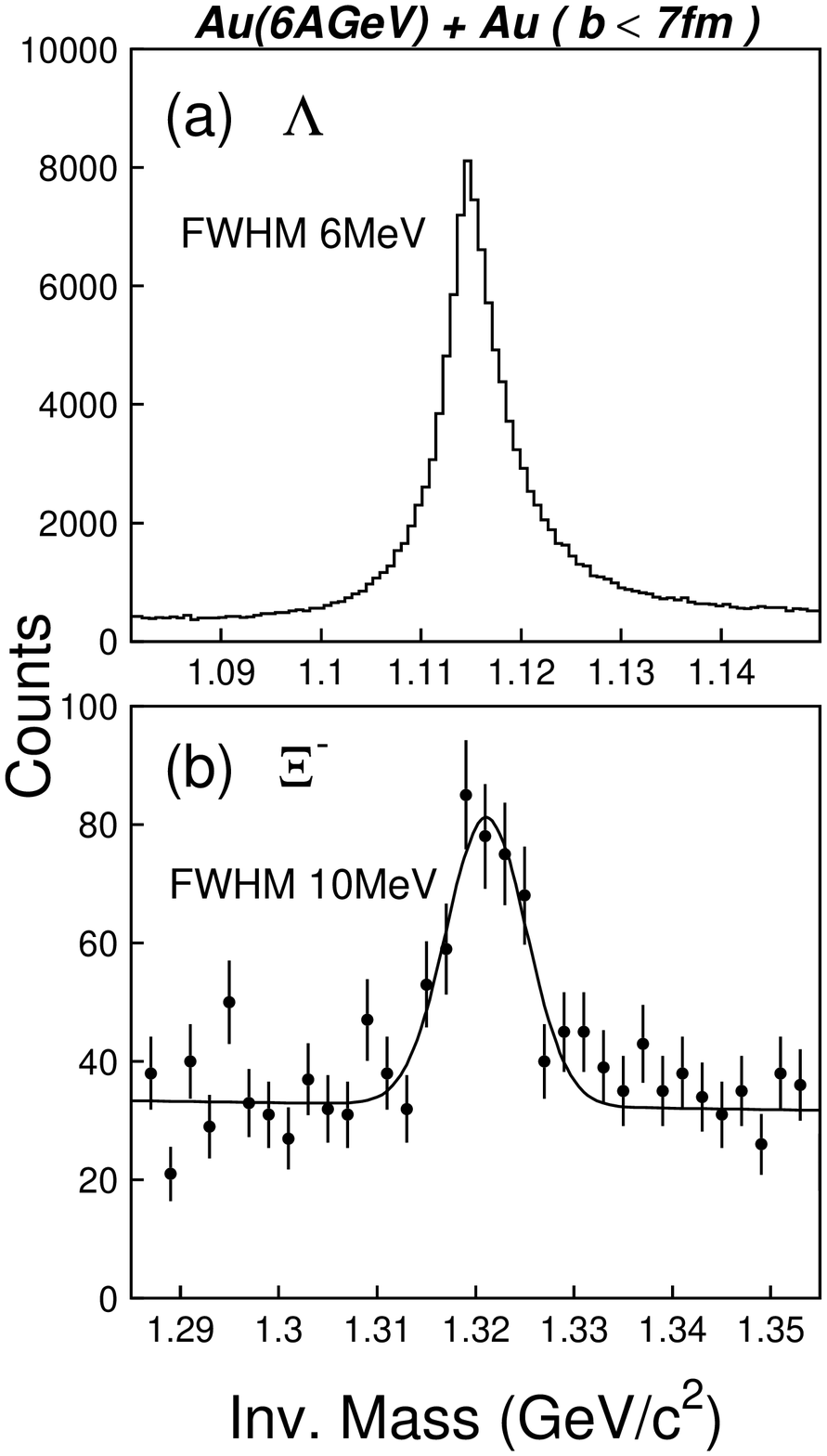}}
\vspace*{0.2in}
\caption{Invariant mass spectrum for (a) $\Lambda$ and (b) $\Xi^-$ in 6
AGeV semi-central Au+Au collisions ($b<$ 7 fm). The curve in panel (b) is a Gaussian fit
to the data points. The Full Widths at Half Maximum for the peaks are
indicated. With these cuts there are  $\sim$~250 $\Xi^-$ counts (net of background).
}
\label{mass_peak}
\end{figure}

	Several studies have demonstrated the viability of vertex and trajectory
reconstruction for V$^0$'s, i.e. from a track pair resulting from two charged
daughter particles \cite{PChung98,Justice98,Justice97}. However, a 
signature for the doubly-strange $\Xi^-$ must be demonstrated by 
finding the two consecutive decays:
$\Xi^- \rightarrow \Lambda \pi^-$ followed by $\Lambda \rightarrow
p \pi^-$. This requires locating three correlated tracks and is
considerably more difficult, especially for low yields near the threshold.
The upper and lower panels of Fig.~\ref{mass_peak} show signature peaks in 
the invariant mass spectra for primary $\Lambda$'s and for the $\Xi^-$ respectively.
The $\Xi^-$ mass peak is well centered at its rest mass of 1.32 GeV.
The width of the distribution for $\Xi^-$ (FWHM) is  10 MeV compared to 6
MeV for the width of the $\Lambda$ as can be expected from the resolution of
the TPC.

	A description for the reconstruction of primary $\Lambda$'s is 
given in Refs.~\cite{PChung98,Justice98}. $\Xi^-$ reconstruction proceeded via a combination 
of two distinct steps. In the first, primary and secondary $\Lambda$-hyperons were reconstructed 
from the daughters of their charged particle decay, $\Lambda \longrightarrow p + \pi^-$ 
(branching ratio $\sim$ 64\%) following essentially the procedure outlined in
Refs.~\cite{PChung98}~and~\cite{Justice98}, except that the training (see below) 
was different for $\Lambda$'s that do not originate directly from the primary vertex 
(secondary $\Lambda$'s).
All TPC tracks in an event were reconstructed followed by the
calculation of an overall event vertex. Thereafter, each $p\pi^-$
pair was considered and their point of closest approach obtained.
Pairs whose member trajectories intersect (with fairly loose criteria such as
decay distance from the event vertex  $>$ 0.5 cm ) at a point
other than the main event vertex were assigned as $\Lambda$ candidates
(primary and secondary) and evaluated to yield an invariant mass and associated momentum.
These $\Lambda$ candidates were then passed to a fully connected
feed-forward multilayered neural network\cite{Justice97} trained to separate ``true''
secondary $\Lambda$'s  from the combinatoric background. The
network was trained from a set consisting of ``true'' secondary $\Lambda$'s
and a set consisting of a combinatoric background. ``True'' secondary
$\Lambda$'s were generated by tagging and embedding simulated $\Lambda$'s
(created from the decay of $\Xi^-$ particles from RQMD\cite{rqmd} calculations) 
in raw data events in a detailed GEANT simulation of the TPC. The combinatoric
background or ``fake'' $\Lambda$'s were generated via a mixed event
procedure in which the daughter particles of the $\Lambda$ ($p \pi^-$) were chosen
from different data events.

	In step 2, the $\Xi^-$ particle was reconstructed from its weak 
decay: $\Xi^- \longrightarrow \Lambda \pi^-$ (branching ratio = 100$\%$; c$\tau =$ 4.9~cm),
using the secondary $\Lambda$'s found in step 1. To do this, each $\Lambda\pi^-$ pair was 
evaluated to assign $\Xi^-$ candidates [and their associated invariant mass and momentum]
which were passed to a fully connected feed forward multilayered neural 
network\cite{Justice97} trained to separate ``true''
$\Xi^-$'s from the combinatoric background. For this latter step, the network was trained from a
set consisting of ``true'' $\Xi^-$'s and a set consisting of a combinatoric
background following a procedure similar to that described earlier for $\Lambda$'s.

	For the four centrality selections (central to peripheral) presented here 
71656, 140706, 154542 and 57236 events were processed to yield 78, 118, 100 and 16
$\Xi^-$'s respectively. The corresponding number of primary $\Lambda$'s are 21526, 
37090, 30973 and 6994 respectively. Relatively good phase space 
coverage was achieved for both the $\Xi^-$'s ($0 < p_T < 1.2$ GeV/c, $-0.5 < y_n < 0.5$)  
and the $\Lambda$'s ($0 < p_T < 1.2$ GeV/c, $-0.7 < y_n < 0.6$). Here, $y_n$ is the 
normalized rapidity. The detection efficiency for $\Xi^-$ ranged 
from $5.4 \pm 0.4 \times10^{-3}$ to $23.5 \pm 0.5 \times 10^{-3}$. Similar efficiencies for 
primary $\Lambda$'s ranged from $2.3 \pm 0.2 \times 10^{-2}$ to $6.8 \pm 0.2 \times 10^{-2}$.

	The neural network optimizes (during training) and then applies 
(during reconstruction) a multidimensional cut on topological variables 
or input neurons\cite{Justice97}, to distinguish true $\Xi^-$'s.  In addition, 
several $\lq\lq$hard cuts" (e.g. $dca_{\Lambda,\pi}$, where $dca$ is the 
distance of closest approach of a ${\Lambda,\pi}$ to the event vertex) 
were employed to investigate and minimize systematic uncertainties. 
Figure \ref{mass_peak} shows the results 
obtained with such restrictions. In general, tighter $\lq\lq$hard cuts" led to 
improvements in the signal-to-background ratio, but with significant reductions in the 
raw yield due to poorer detection efficiencies. Nevertheless, the 
efficiency corrected yields showed essentially no dependence on these cuts. 

	For the different cut conditions there are, of course, different 
efficiencies for track finding and thus for $\Xi^-$ and $\Lambda$ reconstruction. These 
efficiencies have been determined by simulations that
involve embedding  $\Xi^-$ particles from RQMD calculations into events
from the data sample and running this modified data set through the neural
network. Subsequently the ratio of reconstructed $\Xi^-$ particles in the 
real data divided by the simulated efficiency has been studied for a large 
variety of cut conditions. This method allows us to minimize systematic 
errors for the applied hard cut and thus we expect that statistical 
errors predominate. As a further test of this efficiency determination method, 
a model-independent integration over $p_T$ and rapidity was made 
for $\Lambda$'s. The result was in very good agreement with the method 
used here for both $\Lambda$'s and $\Xi^-$'s.

\begin{figure}
\centerline{\epsfysize=3.5in \epsffile{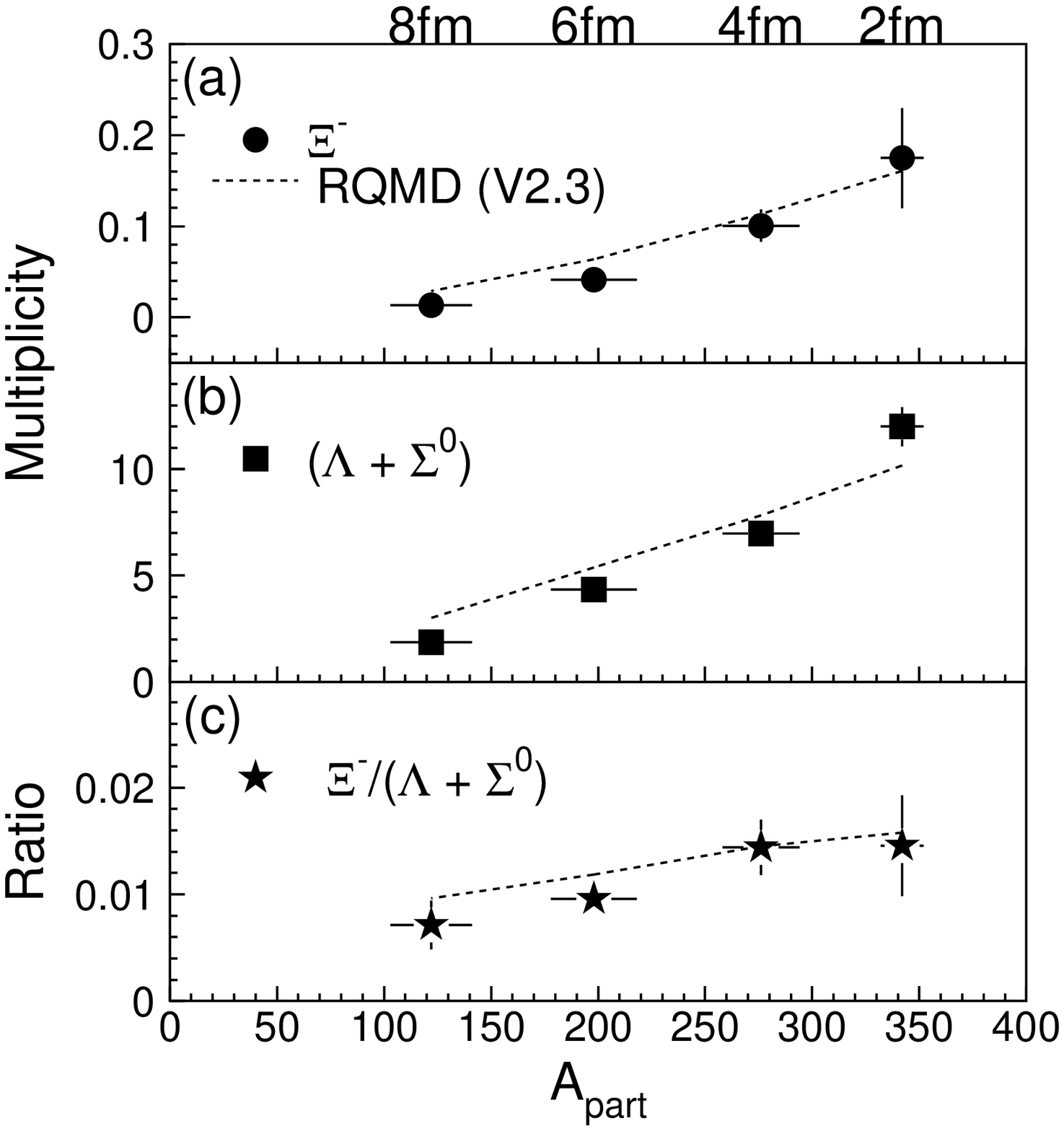}}
\vspace*{0.2in}
\caption{Panels (a) and (b) show the A$_{\rm part}$ dependence of the
multiplicity of $\Xi^-$ (circles) and $(\Lambda + \Sigma^0)$ (squares), respectively,
from 6 AGeV Au + Au collisions. Central collision have A$_{\rm part} \sim 370$. 
Panel (c) shows the corresponding
multiplicity ratio (stars), $\Xi^-$/($\Lambda + \Sigma^0$), as a function of A$_{\rm part}$.
Values from RQMD calculations are indicated via the dotted lines.
}
\label{mult_vs_b}
\end{figure}

	Several centrality cuts have been made in order to examine the 
variation of the $\Xi^-$ multiplicity on impact parameter and the 
number of participating nucleons A$_{\rm part}$. 
For each centrality cut, an impact parameter range was assigned via
its respective fraction of the minimum bias cross section as described 
in Ref.\cite{chu02}. The corresponding values for A$_{\rm part}$ were 
obtained via the Glauber model\cite{glauber}.
Figures \ref{mult_vs_b}a and b show the respective centrality (A$_{\rm part}$) 
dependence of the multiplicity for $\Xi^-$ and ($\Lambda + \Sigma^0$) 
as the $\Lambda$ yield includes feed-down contributions from the $\Sigma^0$. 
The ratio $\Xi^-$/($\Lambda + \Sigma^0$) 
is shown as a function of A$_{\rm part}$ in Fig. \ref{mult_vs_b}c. 
Figures~\ref{mult_vs_b}a and b indicate a relatively strong increase in 
the $\Xi^-$ and ($\Lambda + \Sigma^0$) multiplicities with 
increasing A$_{\rm part}$. However, this dependence is greater for the $\Xi^-$
as can be seen in Fig. \ref{mult_vs_b}c. An extrapolation [with the fit 
function $multiplicity$ = const~$\times$~A${_{\rm part}}^{\alpha}$] to the data 
shown in Fig.~\ref{mult_vs_b}c leads to the ratio 
$\Xi^-$/($\Lambda + \Sigma^0$) of $0.017 \pm 0.005$ 
for most central collisions ($b$=0 fm or A$_{\rm part} \sim $ 370).

\begin{figure}
\centerline{\epsfysize=3.5in \epsffile{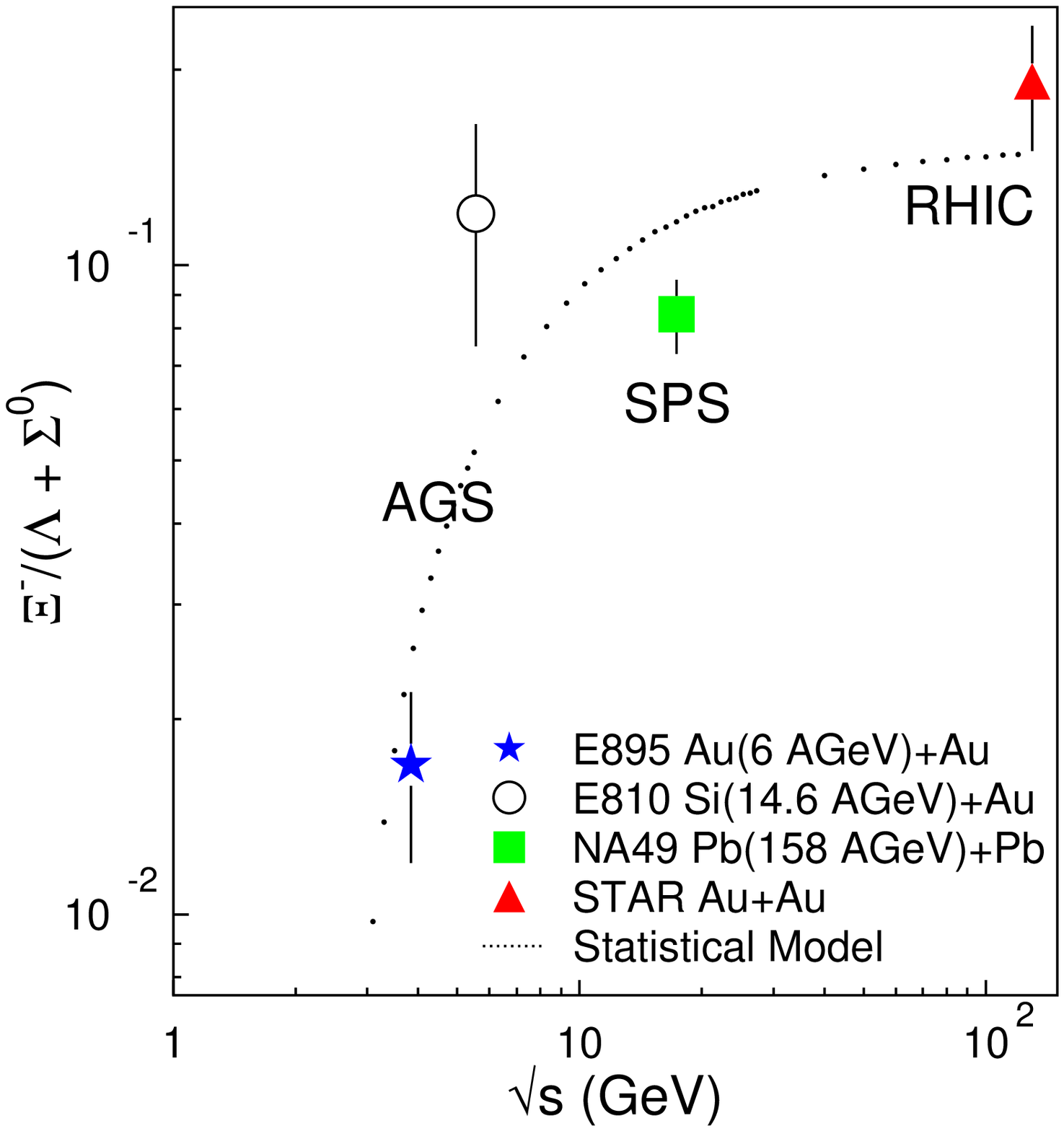}}
\vspace*{-0.2in}
\caption{
The multiplicity ratio of $\Xi^-/(\Lambda + \Sigma^0)$ as a function of
$\sqrt(s)$
Filled symbols correspond to central collisions of Au+Au/Pb+Pb. The datum indicated
by the open circle is for the asymmetric system Si+Au. The dotted line shows results
from the statistical model for central Au+Au collisions.
}
\label{yield_ratio}
\end{figure}

	Figure \ref{mult_vs_b} also shows the results of calculations from the
transport model RQMD\cite{rqmd} as indicated. For the single-strange hyperons 
$\Lambda + \Sigma^0$, the model prediction of an increase in multiplicity with 
increasing centrality follows the trend of the data rather well. A similarly good 
agreement is apparent for the doubly-strange $\Xi^-$ hyperon. 
This rather striking agreement between data and theory is in direct 
contrast to the observation that transport models are unable to describe the 
measured yields for  $\Xi^-$ at SPS energies \cite{Soff}.
Next we compare the ratio $\Xi^-$/($\Lambda + \Sigma^0$) to that obtained at higher
incident energies and to statistical model predictions as shown in Fig.~\ref{yield_ratio}.

	The statistical model makes predictions for this ratio as a function of
$\sqrt s$ \cite{max_strange}. The general freeze-out curve \cite{CR} taken together 
with the values of the chemical freeze-out temperature $T$ and baryochemical 
potential $\mu_B$ as a function
of $\sqrt s$  \cite{max_strange} define all of the particle ratios.
The dotted line in Fig.~\ref{yield_ratio} indicates the ratio
$\Xi^-/(\Lambda + \Sigma^0)$ calculated for central collisions. 
It exhibits a pronounced rise in the AGS/SPS energy regime and then 
flattens off toward the RHIC domain of $\sqrt s >$~100 GeV. 
Parameters for the statistical model are determined from the systematics 
of fits to experimental data, and hence have finite uncertainties. Therefore, the curve 
in Fig. 3 should in principle be replaced by a band. Evaluation of these uncertainties 
as a function of energy is beyond the scope of this work.

	The rise up to SPS energies results from two primary effects.
The first is an increase in temperature, from $\approx$ 50 MeV
(SIS, GSI) up to $\approx$ 170 MeV (SPS/RHIC). 
The second effect originates from the fact that one needs to use a 
canonical description at the lower incident energies. 
For central collisions at SPS energies, the number of multi-strange 
particles is high and a grand-canonical description (global strangeness 
conservation) is sufficient. At lower incident energies, this number 
is much smaller requiring a canonical treatment (local strangeness conservation)
which leads to an additional suppression towards the lower incident
energies \cite{Clemans_01}. Above top SPS energies the temperature is roughly constant, 
while $\mu_B$ decreases. Since the ratio $\Xi^-/(\Lambda + \Sigma^0)$ is 
independent of $\mu_B$, it saturates (neglecting higher resonances which are 
taken into account in the model calculations)to the limit 
$g_{\Xi^-}/(g_{\Lambda} + g_{\Sigma^0}) \; \cdot \; exp[-(E_{\Xi^-} + \mu_S -
E_{\Lambda,\Sigma^0})/T]$ with $g_i$ the degeneracy, $\mu_S$ the strangeness 
chemical potential and $E_i$ the energy of particle $i$. 
Detailed calculations for the centrality dependence are 
not yet available.

	Data for $\Xi^-$ production are indeed scarce. Results obtained 
at SPS from NA49~\cite{NA49} and very recently at RHIC \cite{Castillo} for central
collisions of Pb+Pb/Au+Au are shown as filled symbols in Fig.~\ref{yield_ratio}.
The RHIC point [for midrapidity only] agrees (well) with the statistical model. 
The SPS point for the $4\pi$ integrated yield is below the statistical-model curve.
At the highest AGS energy a data point (open circle) is shown for a Si+Au 
measurement performed at slightly forward rapidities (y =1.4-2.9) ~\cite{Si_Au}. 
This data point exceeds the curve, but is within two
standard deviations even neglecting the systematic uncertainty in the 
statistical model calculations.
The filled star in Fig.~\ref{yield_ratio} indicates our new $4\pi$ integrated
result for central Au+Au collisions at 6 AGeV. This value is in good  
agreement with the statistical model prediction \cite{max_strange}.

This data point is also in good agreement with the predictions of 
the RQMD model (cf. Fig.~\ref{mult_vs_b}c). Thus, we conjecture that hadronic 
processes are able to explain the measured yields. This is in 
agreement with recent UrQMD calculations \cite{reiter2003} which give 
a detailed account of the channels responsible for $\Xi^-$ production. 
Since the statistical model predictions also give a good representation 
of the data, the question is raised as to how far transport model calculations 
are from chemical equilibrium. Additional theoretical work is clearly needed 
to answer this question and to unravel the production mechanism/s for multistrange 
hadrons from energies close to threshold up to the SPS [where current calculations 
fail to reproduce the measured yields by hadronic processes alone]. 

	In conclusion, we have presented the first measurement for the doubly 
strange hyperon $\Xi^-$ in near-threshold Au+Au collisions. The multiplicity 
of the $\Xi^-$ is observed to increase more rapidly with centrality than the 
multiplicity of $\Lambda + \Sigma^0$, but the measured ratio  
$\Xi^- / (\Lambda+\Sigma^0$) for central collisions agrees well with both 
statistical and transport model predictions. This agreement may be an indication
that despite the rather short reaction times at AGS energies, the available 
phase space for $\Xi^-$ is essentially filled. 
%

        This work was supported in part by the U.S. Department
of Energy under grants DE-FG02-87ER40331, DE-FG02-89ER40531,
DE-FG02-88ER40408, DE-FG02-87ER40324, and contract
DE-AC03-76SF00098; by the US  National
Science Foundation under Grants  PHY-98-04672, PHY-9722653,
PHY-96-05207,PHY-9601271, and INT-9225096; by the University of
Auckland Research Committee, NZ/USA Cooperative Science Programme
CSP 95/33; and by the National Natural Science Foundation of P.R. China
under grant 19875012.


\end{multicols}
\end{document}